\documentclass[12pt,authoryear]{elsarticle}
\usepackage{graphicx}
\usepackage{amssymb}
\usepackage{amsmath}
\usepackage{amsthm}

\journal{Information and Software Technology}

\usepackage{microtype}
\usepackage[svgnames]{xcolor}
\usepackage{algorithm}
\usepackage{algpseudocode}
\usepackage{url}
\usepackage{booktabs}
\usepackage{tabularx}
\usepackage{thmtools}
\usepackage{multirow}
\usepackage{makecell}
\usepackage{adjustbox}

\usepackage{tikz}
\usetikzlibrary{arrows.meta, positioning, backgrounds, bending}

\pgfdeclarelayer{background}
\pgfsetlayers{background,main}

\usepackage{enumitem}

\usepackage{pifont}
\newcommand{\cmark}{\ding{51}}%
\newcommand{\xmark}{\ding{55}}%

\usepackage{subcaption}

\declaretheorem[style=plain]{definition}

\declaretheoremstyle[
headfont=\bfseries,
postheadhook = {\hspace*{\parindent}},
spaceabove = 0.5cm, 
spacebelow = 0.5cm]{relation}

\usepackage[theorems,breakable]{tcolorbox}
\newtcbtheorem{relation} %
  {Metamorphic Relation}
  { %
    breakable,
    colback=white, %
    colframe=LightGray,    %
    coltitle=Black,
    fonttitle=\bfseries   %
  }
  {mr} %

\usepackage{xspace}
\newcommand{\mrmirror}{\(\text{MR}_{Mirror}\)\xspace}
\newcommand{\mrscale}{\(\text{MR}_{Scale}\)\xspace}
\newcommand{\mrrot}{\(\text{MR}_{Rot}\)\xspace}
\newcommand{\mrcc}{\(\text{MR}_{ClsChg}\)\xspace}
\newcommand{\mrobs}{\(\text{MR}_{Obs}\)\xspace} %

\newcommand{\tool}{TrajTest\xspace}

\begin{document}

\begin{frontmatter}

\title{Metamorphic Testing of Multimodal Human Trajectory Prediction}

\author[srl]{Helge Spieker\corref{cor1}} %
\ead{helge@simula.no}
\cortext[cor1]{Corresponding author}

\affiliation[srl]{organization={Simula Research Laboratory},
            city={Oslo},
            country={Norway}}

\author[lisn]{Nadjib Lazaar} %
\ead{lazaar@lisn.fr}

\author[srl]{Arnaud Gotlieb} %
\ead{arnaud@simula.no}

\author[srl]{Nassim Belmecheri} %
\ead{nassim@simula.no}

\affiliation[lisn]{organization={LISN, Université Paris-Saclay}, 
            city={Saclay},
            country={France}}

\begin{abstract}
Context: Predicting human trajectories is crucial for the safety and reliability of autonomous systems, such as automated vehicles and mobile robots. However, rigorously testing the underlying multimodal Human Trajectory Prediction (HTP) models, which typically use multiple input sources (e.g., trajectory history and environment maps) and produce stochastic outputs (multiple possible future paths), presents significant challenges. The primary difficulty lies in the absence of a definitive test oracle, as numerous future trajectories might be plausible for any given scenario.\\
Objectives: This research presents the application of Metamorphic Testing (MT) as a systematic methodology for testing multimodal HTP systems. We address the oracle problem through metamorphic relations (MRs) adapted for the complexities and stochastic nature of HTP.\\
Methods: We present five MRs, targeting transformations of both historical trajectory data and semantic segmentation maps used as an environmental context. These MRs encompass: 1) label-preserving geometric transformations (mirroring, rotation, rescaling) applied to both trajectory and map inputs, where outputs are expected to transform correspondingly. 2) Map-altering transformations (changing semantic class labels, introducing obstacles) with predictable changes in trajectory distributions. We propose probabilistic violation criteria based on distance metrics between probability distributions, such as the Wasserstein or Hellinger distance.\\
Results: The empirical evaluation on a popular HTP model called Y-net demonstrated the feasibility and effectiveness of \tool on this dataset. For label-preserving MRs, the oracle-less Wasserstein violation criterion identified violations with statistically significant agreement relative to ground-truth-dependent metrics, confirming its utility. Map-altering MRs successfully triggered expected changes, such as statistically significant decreases in path probabilities over areas made less walkable or obstacle avoidance.\\
Conclusion: This study introduces \tool, a MT framework for the oracle-less testing of multimodal, stochastic HTP systems. It allows for assessment of model robustness against input transformations and contextual changes without reliance on ground-truth trajectories.
\end{abstract}

\begin{keyword}
Software Testing \sep Metamorphic Testing \sep Human Trajectory Prediction \sep Machine Learning \sep Stochastic Systems
\end{keyword}

\end{frontmatter}

\section{Introduction}
Building safe and reliable autonomous systems requires an accurate prediction of human trajectories. This is true not only for automated vehicles that must plan their trajectories to avoid hitting any pedestrian \citep{Levinson2011} but also in surveillance systems~\citep{VV2005}, in robotics~\citep{Foka2010}, or in planning~\citep{Luo2018}. Human trajectory prediction aims to predict future possible paths taken by individual humans using their past trajectories.
In automated driving, the focus is on predicting short-term, i.e., a few seconds, future trajectories of vulnerable road users, such as pedestrians, cyclists, and disabled people.
Due to their difference in movement patterns compared to vehicles and their inherent vulnerability to urban traffic, predicting the trajectories of these entities is a distinct and critical task.
Human trajectory prediction is an active research area and current methods have achieved strong results~\citep{li2022graph,xu2022groupnet,Bae_2022_CVPR,duan2022complementary,Shi_2021_CVPR,Dendorfer_2021_ICCV,Mohamed_2020_CVPR,mangalam2020not}.
Despite these recent advances, ensuring the robustness, accuracy, and reliability of these prediction models is still a challenge, as detailed below. 
It is crucial to rigorously test these models to identify potential flaws, evaluate their performance, and ensure their safety for practical integration into autonomous systems~\citep{Uhlemann2024}; with adversarial attacks being the main robustness testing technique for HTP in the existing literature~\citep{Zhang_2022_CVPR,cao2022advdo,pmlr-v205-cao23a,Zheng_2023_WACV,pmlr-v211-tan23a,jiao2022semi}.
Here, a key focus is on the perturbation of the historical trajectory to maximize prediction errors (and ultimately to make the model more robust), but less emphasis on a structured manipulation of the different input sources of a HTP system.

Given that human trajectory prediction models are machine learning systems and operate stochastically, testing not only serves to detect bugs, but also to evaluate and measure the model performance.
While human trajectory prediction datasets usually contain a ground-truth of trajectories, it is noteworthy that many alternative trajectories could have been similarly realistic. Thus, implementing a broader evaluation scheme, beyond simply measuring the distance to a singular ground-truth trajectory, would provide a richer understanding of the robustness and generalizability of the method used. Current trajectory prediction models are multi-source, taking into account multiple sources of information~\citep{Fu2024}, including the past trajectory of pedestrians, the environmental map, and possibly the interactions between humans. 

Metamorphic testing (MT) is a relevant approach for testing programs that do not have oracles available. 
As such, this approach is particularly relevant for validating AI systems that embed trained models. Introduced in \citep{Chen1998}, MT replaces traditional oracle checking with {\it metamorphic relations (MRs)}, which are necessary but not sufficient properties that must be satisfied by the software or model under test. By assessing the results of multiple program executions \citep{Segura2016,Chen2018}, MRs can automatically detect bugs. MRs can be used to generate the so-called {\it follow-up test cases} \citep{Chen1998, Chen21} to check the specific relations among the results of the software or model under test. MT has been successfully deployed to test a variety of complex software systems, including ML models \citep{Xie2009, Xie2011,Xu2018,Spieker2020,Xiao22,duran2025metamorphic}, scientific software \citep{Yoo2010,Kanewala2016}, or virtual reality applications \citep{Alves23}. Interestingly, MT has received considerable attention in the field of automated driving~\citep{Zhou2019, DLZ21, Deng2022, AVS23} and stochastic systems~\citep{guderlei2007statistical,Yoo2010,Chen2018}, of which HTP are a special case, but to the best of our knowledge, it has not yet been applied to test human trajectory prediction models.

In this article, we argue and demonstrate that MT can be successfully applied to multimodal human trajectory prediction by being a pragmatic approach to address the complexities and non-determinism of these models. Using traditional testing is limited to the available ground-truth data and the challenge of obtaining accurate datasets for a broad variety of scenarios and situations. 
MT enhances robustness by generating additional, diverse test data to identify edge cases and subtle errors by validating MRs. 
We apply MT for HTP in the sense of traditional testing, but also to expand the evaluation setting by providing a more diverse view of the robustness of the models under input transformations without requiring additional involvement in data collection and labelling.
The contributions of this article are threefold:
\begin{enumerate}
\item We introduce five MRs dedicated to HTP testing that are based on the modification of the different inputs required for HTP models, namely the original image, the segmentation map, and the historical trajectory. We re-visit MRs from the broad context of image analysis and processing models and tuning them for the novel specific case of HTP testing. Also, we evaluated very different MRs for HTP testing to broaden the scope of the proposed approach; 
\item We formalize MT for HTP by using and comparing three novel and different violation criteria that are used to determine the violation of MRs. These criteria, respectively called the probabilistic, Wasserstein and Hellinger violation criterion, capture the non-deterministic nature of HTP predictions and deals with absence of a unique ground truth for the predictions;
\item We perform an illustrative experimental evaluation on popular trajectory prediction systems, namely Y-net~\citep{Mangalam_2021_ICCV} on the Stanford Drone Dataset (SDD)~\citep{robicquet2016learning} and the intersection drone dataset (inD)~\citep{Bock2020}.
\end{enumerate}

An earlier version of this study appeared at the 9th ACM International Workshop on Metamorphic Testing~\citep{Spieker2024}. 
We extend the previous work through (a) a new class of metamorphic relations, oriented on the manipulation of the semantic map, (b) the Hellinger violation criterion as an alternative to the previously introduced Wasserstein violation criterion, (c) extended experiments including a second prediction setting on the SDD dataset, and the inD intersections drone dataset as an additional dataset, and, finally, (d) more thorough and detailed discussions of the methodology.

The rest of the paper is organized as follows: Section~\ref{sec:background} introduces some background information on MT and HTP. 
Section~\ref{sec:rel_work} presents the SotA in HTP testing. 
Section~\ref{sec:framework} introduces our MT for multimodal HTP testing framework. 
Section~\ref{sec:evaluation} presents our empirical evaluation.
Eventually, Section~\ref{sec:conclusion} concludes the article and draws some perspectives.

\section{Background}\label{sec:background}

\subsection{Metamorphic Testing (MT)}
\label{sec:mt}
Metamorphic Testing aims at testing programs that do not have available oracles. Such programs include supervised machine learning models that generalize their predictions after being trained on a set of labelled instances \citep{Zhang2020}. 
The exact behaviour of these models largely depends on the data sets used for the training, and their predictions are usually affected by possible data over- or under- fitting and uncertainties in their generalization abilities. 
MT has been used to test simple classifiers \citep{Murphy2008,Xie2011}, deep learning models~\citep{Ding2017a}, machine translation \citep{sun2018metamorphic}, chess engines~\citep{DBLP:journals/infsof/MartinKMA25}, AI planning~\citep{mazouni2025mutation}, automated driving~\citep{DBLP:journals/tse/DengZZLLKC23}, object detection and classification~\citep{Spieker2020}, and human pose estimation~\citep{duran2025metamorphic}. 
MT relies on the availability of Metamorphic Relations (MRs)~\citep{Chen1998, Chen2018}:
\begin{definition}[Metamorphic Relations]\label{DefMR}
Let $P$ be a program under test, $x$ and $y$ two test inputs for $P$, then an MR for $P$ is expressed as a relation $\forall x, \forall y, r_i(x,y) \implies r_o(P(x), P(y))$ where $P(x)$ (resp. $P(y)$) denotes the execution of $P$ on $x$ (resp. $y$) and $r_i$ and $r_o$ correspond to relations with the inputs and outputs of $P$. 
\end{definition}
It should be noted that MRs are necessary (but not sufficient) properties to ensure the correctness of $P$ w.r.t. its specification. Formally speaking, $r_i(x,y) \wedge \neg r_o(P(x),P(y)) \implies \neg correct(P)$. MRs are convenient properties for generating test cases. 
Let $r_i(x,y) \implies r_o(P(x), P(y))$ be an MR for $P$, then if there exists a transformation $t$ (possibly non-deterministic) such that $y=t(x)$ and $t \subseteq r_i$, then it becomes possible to generate a sequence of test cases from $x$, namely $<x,t(x), t(t(x)), ...>$ which all have to fulfil the MR for $P$. Indeed, if $t^{i}$ denotes $i$ successive applications of $t$, it is trivial to see that $\forall i, r_i(t^{i}(x), t^{i+1}(x))$ holds as $t \subseteq r_i$ and thus $r_o(P(t^{i}(x)), P(t^{i+1}(x)))$ holds as $P$ must satisfy the given MR.
\begin{definition}[Follow-up Test Case]\label{DefFTC}
For a given MR and any $t \subseteq r_i$, $t(x)$ is called a {\it follow-up test case} of $x$ for that MR.
\end{definition}
As noted in Segura's survey~\citep{Segura2016}, many MRs can usually be identified to test a program. Then, a key difficulty in MT is finding MRs and determining which ones have the greatest fault-revealing capabilities. 

One advantage of applying MT and MRs is that they can be highly specific to the system-under-test, but often they can be designed general to be transferable between systems and domains, as long the type of input source remains the same.
For example, when testing systems with image-based inputs, standard geometric transformations can be applied independent of the exact system, e.g., as has been done in the literature using standard transformations like mirroring or rotation~\citep{Spieker2020,duran2025metamorphic,DBLP:journals/jss/XuTFBZC21}. 

\subsection{Human Trajectory Prediction (HTP)}

HTP has received considerable attention in the last two decades due to the emergence of sensor-based crowd surveillance, service robotics and automated driving applications.
HTP explores the capabilities of AI models to predict human paths in various environments. In automated driving, HTP plays a pivotal role between the perception capabilities of automated vehicles and the decision-making modules~\citep{Fu2024}. 
HTP techniques differ by i) the input data nature, that may or may not account for vehicle-to-pedestrian and social interactions between pedestrians; ii) the diversity of AI methods employed, ranging from traditional methods based on rules to the most advanced deep learning architectures such as transformers; iii) output characteristics aiming to improve forecast accuracy.

Classic methods based on pedestrian dynamics, such as velocity and acceleration or Bayesian inference~\citep{Bera2016, lee2018} that learn the motion patterns of pedestrians, have been quickly overcome by methods based on deep learning. By learning representations from data, these methods can capture complex and unexpected interactions between humans, vehicles, and other entities from a given scene. In this context, trained models have evolved to handle both {\it unimodal} and {\it multimodal} outcomes. 
Unimodal models focus on predicting a single probable future path, as seen in methods such as Social Forces~\citep{helbing1995social} and Social LSTM~\citep{Alahi2016SocialLH}. In contrast, multimodal models address the uncertainty in prediction by providing multiple potential future paths. Generative approaches such as DESIRE~\citep{lee2017desire}, Trajectron++~\citep{salzmann2020trajectron++}, and Introvert~\citep{shafiee2021introvert} utilize learned latent space variables to generate stochastic outcomes for future predictions. Furthermore, models like those proposed by Liang et al., Mangalam et al., and Zhao et al. employ spatial probability estimates to capture multimodality through probability maps \citep{liang2020simaug,mangalam2020disentangling,zhao2021tnt}.

Depending on the prediction model, different input sources are used to make a prediction, such as the human pose and gaze of other pedestrians in the scene~\citep{Fu2024}. 
These input signals can reveal the immediate intentions of the individual and the potential interactions that can influence the trajectory of the individual. 
RGB, radar, or Lidar images of the scene, e.g., from a drone, can offer a comprehensive view of the environment. 
Semantic scenes and location data can provide context, thus enhancing the accuracy of the prediction.

\begin{definition}[Multimodal HTP]\label{defHTP}
Given historical information, the objective of the model is to predict the distribution of a human trajectory for future $T$ timesteps. 
This can be achieved by generating a multimodal probability distribution over plausible future trajectories, conditioned on the map and history.
From this distribution $P_{map}$, a set of K distinct future trajectories $\{Y^{(1)}, Y^{(2)}, \dots, Y^{(K)}\}$ can be sampled to represent a diverse range of likely outcomes.

Formally, the model learns the parameters $\theta$ of the probability function $P_{\theta}(Y|X, M)$ that defines $P_{map}$. 
Here, $X$ represents the trajectory history, $M$ represents the map or information about the environment and $Y$ represents a specific predicted trajectory.
For a given agent $i$, the model's input is the historical information over the past $n$ timesteps, denoted $X_i = (X_{t-n+1}^i, X_{t-n+2}^i, ..., X_t^i)$ and the contextual information $M$.
Each sampled future trajectory $Y^{(k)}$ consists of predicted positions for the next $T$ timesteps from the current time $t$, defined as $Y^{(k)} = (Y_{t+1}^{(k)}, Y_{t+2}^{(k)}, ..., Y_{t+T}^{(k)})$.

\end{definition}
We note that the existence of $P_{map}$ is not a mandatory artifact of every HTP system, but expected for some MRs discussed in this paper.

\section{Metamorphic Testing of Multimodal HTP}\label{sec:framework}

\begin{figure}[t]
  \centering
  \includegraphics[width=\textwidth]{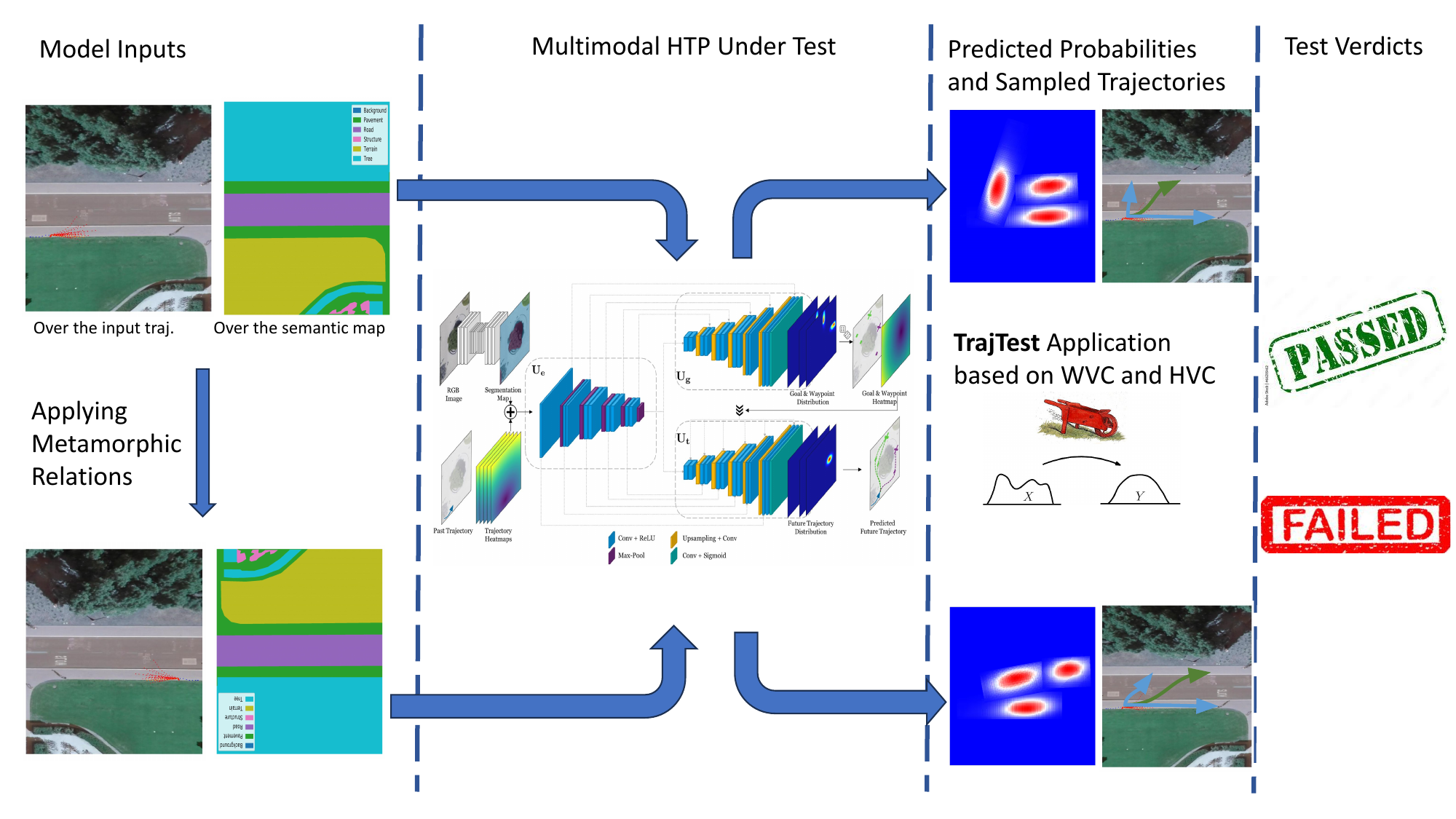}
  \caption{\tool{}: Metamorphic Testing for multimodal HTP}
  \label{fig:diag}
\end{figure}

We present an MT method for multimodal HTP, designed for handling stochastic prediction output. Our framework, called \tool, is illustrated in Figure~\ref{fig:diag}. In the figure, the model under test in the centre is launched multiple times with inputs modified by the proposed metamorphic relations. Current HTP models expect as input the previous trajectory of the human plus additional information~\citep{Fu2024}. The comparison between predicted trajectories is performed using specific violation criteria, namely the Waserstein Violation Criterion (WVC) and the Hellinger Violation Criterion (HVC), which will be introduced in Section~\ref{sec:pvc}. 

\subsection{Testing Multimodal Human Trajectory Prediction Systems}\label{sec:htpsys}

In this work, we consider HTP models where the additional information is a visualization of the scene from a bird-eye view (BEV), i.e., from a position located above the automated vehicle.
This is one common setup in the HTP literature~\citep{mangalam2020not,Mangalam_2021_ICCV,luo2023gsgformer,robicquet2016learning}, although not the only one, depending on the context of the HTP systems and available sensors.
BEV-based HTP models usually compute a \textit{segmentation map} of the environment of the scene.
A segmentation map for automated driving labels different zones within the vehicle's environment. This includes delineating drivable surfaces, such as roads and pedestrian paths, as well as identifying static elements such as buildings, trees, and traffic signs. These elements allow automated vehicle decision-making modules to make an informed choice related to path planning, obstacle avoidance, and safe navigation.

\begin{figure}[t]
    \centering
    \begin{subfigure}[t]{0.48\columnwidth}
        \centering
        \includegraphics[width=\textwidth]{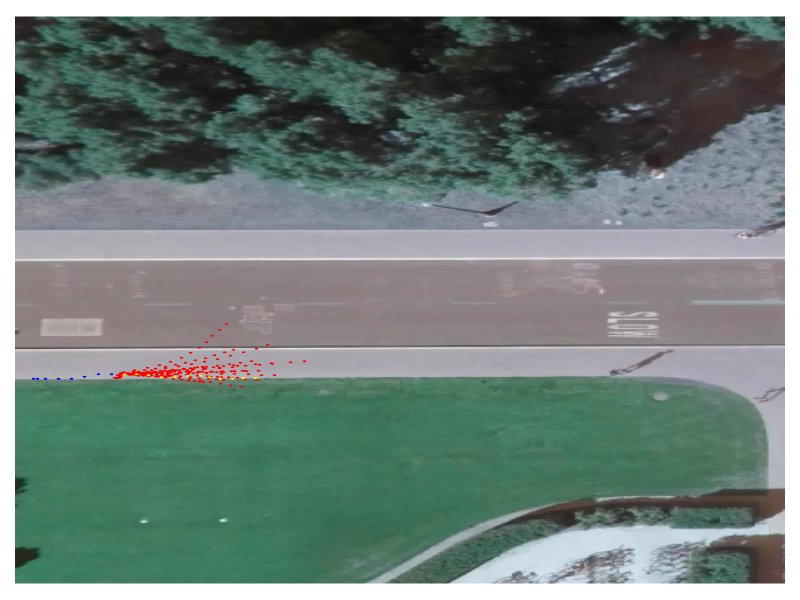}
        \caption{Inputs and Outputs. Blue is the past history, red is the predicted trajectories, yellow is the ground-truth trajectory.}
        \label{fig:input_output}
    \end{subfigure}
    \hfill
    \begin{subfigure}[t]{0.48\columnwidth}
        \centering
        \includegraphics[width=\textwidth]{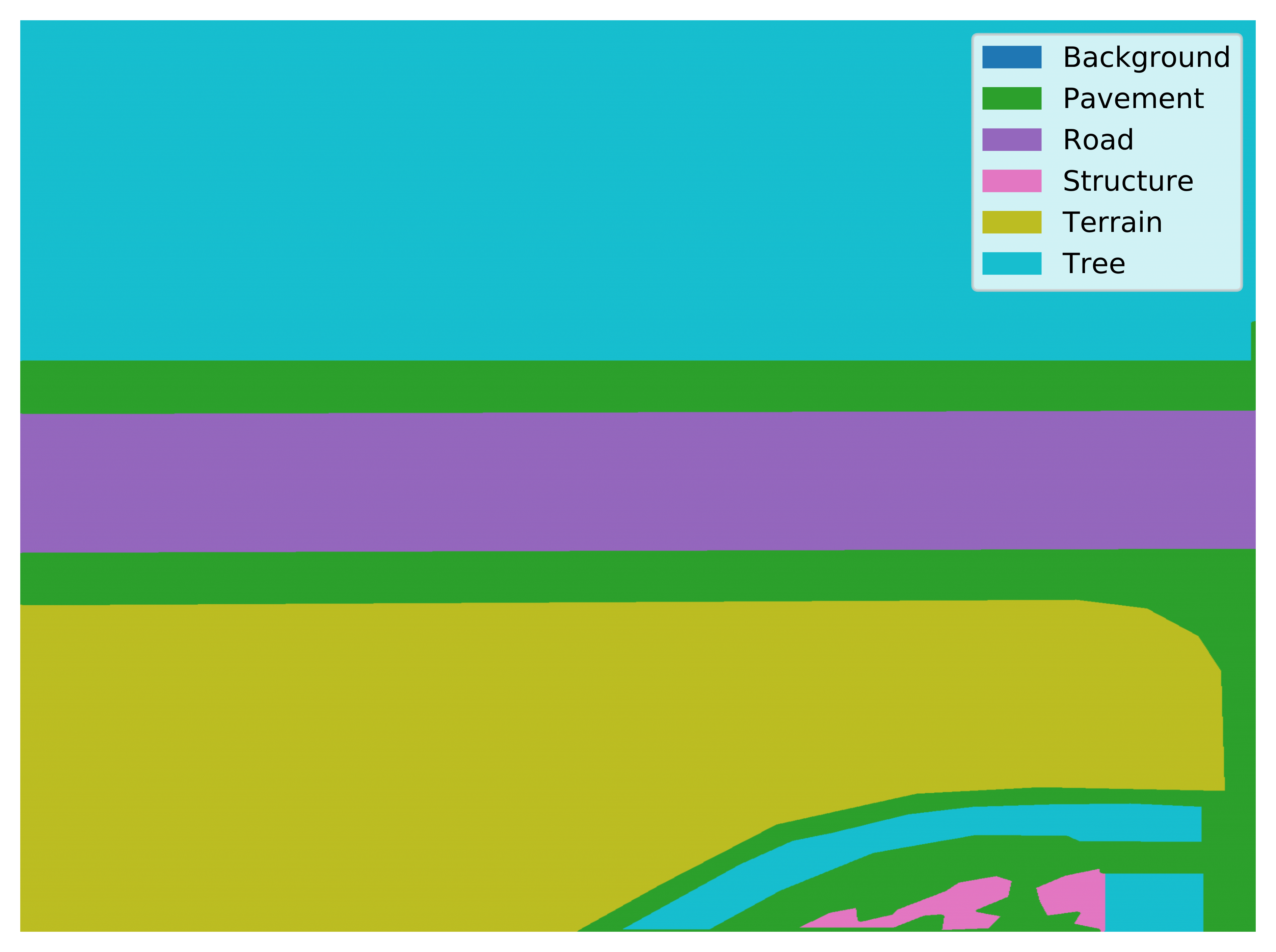}
        \caption{Image Segmentation Output. Six classes can be annotated.}
        \label{fig:segmentation}
    \end{subfigure}
    \caption{Inputs and Outputs of Human Trajectory Prediction. Data shows the little\_1 scene from the Stanford Drone Dataset~\citep{robicquet2016learning}.}
    \label{fig:htp_data}
\end{figure}

Both the historical trajectory and the segmentation map can be modified by metamorphic relations.
The historical trajectory can be manipulated directly, as it is a sequence of 2D Cartesian coordinates $(x,y)$ in BEV.
Manipulating the image input directly is more difficult to do automatically and runs the risk of introducing unrealistic artifacts that hinder the testing process, even with modern generative ML models.
For this reason, we manipulate the input on the level of the segmented image, i.e., after the first input processing step.
This has the disadvantage that it excludes the segmentation model from the test process, which then needs to be tested separately, but makes the overall test setup for HTP testing more approachable and easier to handle.
Figure~\ref{fig:htp_data} visualizes the main inputs and outputs of an HTP model.
The left side shows the original RGB image, the input trajectory (blue) and a set of sampled output trajectories (red).
The right side shows the corresponding segmentation map of the RGB image with color-coded areas. 
In this example, five different area types plus a background class are distinguished, which is common in the literature~\citep{Mangalam_2021_ICCV,luo2023gsgformer}, but other class structures are possible.

In addition to the predicted trajectories, some HTP systems, including Y-net~\citep{Mangalam_2021_ICCV}, first predict the probability that each point on the map is the goal of the pedestrian. From this probability map, they heuristically sample (intermediate) waypoints, expected final goals, and eventually the predicted trajectories.
We consider this intermediate probability map an additional artifact to be included in the testing process. 
It provides a larger overview of the SUT's assessment of the overall environment and context than the distribution of trajectories, and can serve as an indication for fault localization in the HTP pipeline, i.e., whether the fault occurs during the sampling phase or in the model before.

In the following, we first introduce several violations criteria for follow-up test cases. By violation criteria, we mean how to determine that two model outputs are in contradiction to each other. This is a crucial step to evaluate whether an MR running multiple times the model is violated or not. Then, we introduce the considered MRs of our framework, and we bring everything together into the overall MT process for multimodal HTP.

\subsection{Probabilistic Violation Criterion}\label{sec:pvc}
In many MRs, the violation criterion is a basic comparison, for example, a violation occurs if the result of the follow-up test case is $\{=,\neq,\leq,\geq,<,>\}$ than the result of the source test case.
However, the HTP model is a stochastic system and returns a probability distribution, i.e., the final distribution of future predicted trajectories and intermediate outputs such as the probability map of potential waypoints (see Section~\ref{sec:htpsys}).
Hence, a basic comparison test is not suitable, and we need different violation criteria.

We propose a general novel \textit{probabilistic violation criterion} for the detection of faults in label-preserving MRs in multimodal HTP based on the comparison of the output probability distributions from the source and follow-up test case.

\begin{definition}[Probabilistic Violation Criterion - PVC]\label{DefPVC}
Given a selected distance function ($d$) between probability distributions over a metric space and $\delta >0$, then $PVC^d_{\delta}$ compares the distance between the outputs of the system-under-test $SUT$ from the source and the follow-up test cases $S$ and $F$ and ensures the following:
\begin{equation}
    PVC^d_{\delta}(S, F) = d(SUT(S), SUT(F)) < \delta
\end{equation}
\end{definition}

The arbitrarily $\delta$ threshold is introduced to determine when the difference in predicted trajectories is significant enough to indicate a violation. Simply comparing the trajectories from the original and modified test cases is not reliable on its own.  Because it is difficult to manually set this threshold for every situation, we instead use a statistical method. This involves generating multiple predicted trajectories for the original test case, and then calculating the average and spread (standard deviation) of the differences between them.
For the follow-up test case, a violation occurs if a z-test reports a significant ($\text{p-value} \leq threshold$) difference. The p-value threshold can be adjusted to the MR or kept at the common value of $0.05$.

Although both the predicted trajectories and the probability maps generated by the system represent probability distributions, they are fundamentally different in how they are structured. Due to these structural differences, we need to develop slightly different ways to determine if our rule (the ``violation criterion'') has been broken, taking into account the unique characteristics of each type of output. Our initial focus was on comparing two sets of predicted trajectories, which are the final results of the system, to decide whether they are similar enough or significantly different, indicating a violation of our rule. However, we also want to compare the probability maps themselves, as these maps are the foundation from which the final trajectories are derived.

Following this general motivation, we select two distinct distance definitions for $d$ in Def.~\ref{DefPVC} to account for the two situations, namely the \textit{Wasserstein distance} for distributions of trajectories and the \textit{Hellinger distance} for probability maps as specific instantiations of the $PVC^d_{\delta}$ violation criterion.
The Wasserstein distance is designed for distributions in a metric space, such as trajectories, whereas the Hellinger distance is designed to measure the general similarity between probability distributions. We also discuss a violation criterion based on a statistical hypothesis test that can be applied to the probability map, similar to the HVC, but in non-label-preserving MRs where we do expect the SUT to return different results.

\paragraph{Wasserstein Distance}

We propose the \textit{Wasserstein Violation Criterion} for the detection of faults in label-preserving MRs in HTP.
The WVC approaches the comparison of the two distributions as an optimal transport problem~\citep{peyre2019computational}, that is, it determines the minimal cost to transform one distribution into the other.
Specifically, we compare the trajectory distribution using the \textit{Wasserstein} $W_2$ distance, where the cost considers the squared distance.

Informally, the Wasserstein distance is based on a matching between the sampled trajectories in each set, where the overall distance between the matches is minimal. It is described as the minimal cost to transform one probability distribution into the other, and also referred to as \textit{earth mover distance}~\citep{rubner2000earth}, which visualizes the optimal transport concept for two piles of earth that represent two distributions and should be compared by moving as little earth as possible.
For HTP, the Wasserstein distance is the minimum distance from the trajectories in one distribution to the other, where each trajectory is assigned to exactly one other trajectory. The more similar the two trajectory distributions, the smaller the Wasserstein distance. Formally speaking, the criterion is defined as the square root of the infimum on all possible joint distributions (transport plans) \(\gamma\) that have P and Q as marginals:

\begin{definition}[Wasserstein Distance]\label{DefW}
Let $P = SUT(S)$ (resp. $Q = SUT(F)$) be the predicted trajectory distributions of the source test case $S$ (resp. follow-up test case $F$), let \(\prod(P, Q)\) be the set of all such joint distributions \(\gamma\), and \(\mathbb{E}[...]\) denotes the expected squared distance \(\parallel x - y \parallel_2\) between pairs \((x, y)\) drawn according to the optimal transport plan \(\gamma\) ~\citep{weng2019gan}, then,
\begin{equation}
W_2(P, Q) = \inf_{\gamma \sim \prod(P, Q)} \mathbb{E}_{(x, y) \sim \gamma}[\parallel x - y \parallel_2]
\end{equation}
\end{definition}

The Wasserstein distance is commonly solved as an optimal transportation problem~\citep{peyre2019computational} using specific solvers and methods.
In our work, we rely on the Sinkhorn method~\citep{DBLP:conf/nips/Cuturi13} as implemented in the Python Optimal Transport (POT) library~\citep{flamary2021pot}. 

\paragraph{Hellinger Violation Criterion}
Similarly, we consider the Hellinger distance $H$ \citep{hellinger1909neue} that can compare two discrete probability distributions maps $P = (p_1, p_2, \dots, p_k)$ and $Q = (q_1, q_2, \dots, q_k)$:

\begin{definition}[Hellinger Distance]\label{DefH}
Let $P = SUT_{map}(S)$ (resp. $Q = SUT_{map}(F)$) be the probability distribution map for the source test case $S$ (resp. follow-up test case $F$), then
\begin{equation}
    H_2(P, Q) = \frac{1}{\sqrt{2}} \sqrt{\sum_{i=1}^{k} (\sqrt{p_i} - \sqrt{q_i})^2}= \parallel \sqrt{P} - \sqrt{Q} \parallel_2
\end{equation}
\end{definition}

\paragraph{Hypothesis Testing Criterion}

Given the shared dimensionality of the probability maps, the violation criterion can be rigorously defined by the cell-wise probability differences. We formally termed this as the \textit{Hypothesis Testing Criterion (HTC)}. HTC employs a statistical hypothesis test, such as the Wilcoxon signed-rank test~\citep{wilcoxon1945}, to evaluate whether the probability distributions of the follow-up test case exhibit a statistically significant difference compared to the source test case. Although this interpretation simplifies the inherent meaning of the probability values, it provides a straightforward methodology that is particularly adaptable to localized analyses within the probability map. 
Specifically, when modifications are made to segments of the semantic map, the hypothesis test can be selectively applied to the corresponding subsets of the probability map, enabling targeted validation of probability shifts within those regions. A violation of the metamorphic relation is identified when the hypothesis test yields a statistically significant difference between the two probability distributions. 
Furthermore, the test can be configured as two-sided, to detect any significant deviation, or one-sided, to specifically assess whether the follow-up probabilities are significantly less or greater. 
A comparable violation criterion using a two-sided statistical hypothesis test was previously implemented by \citet{Yoo2010} for the comparative analysis of the results of stochastic optimization.

For the context of our work, we define the HTC as follows:
\begin{definition}[Hypothesis Testing Criterion]
    The Hypothesis Testing Criterion (HTC) is a boolean function that determines if a metamorphic relation is violated. 
    It is parametrized by the two probability maps ($P = SUT_{map}(S), Q = SUT_{map}(F)$), a region of interest (\(R\), a significance level $\alpha$, and the alternative hypothesis ($alternative \in \{\text{two-sided}, \text{greater}, \text{less}\}$). 
    A violation is detected if the p-value returned by a statistical test $\mathcal{T}$ is less than the significance level $\alpha$:
    \begin{equation}
        HTC(P, Q, R, \alpha, alternative) = 
        \begin{cases}
            \text{True} & \text{if } \mathcal{T}(P(R), Q(R), alternative) < \alpha \\
            \text{False} & \text{otherwise}
        \end{cases}
    \end{equation}
    where $P(R)$ (resp. $Q(R)$) is the probability map for the area of interest $R$.
\end{definition}

\subsection{MRs for Multimodal HTP}

We introduce a set of MRs to transform source test cases into follow-up test cases divided into two separate groups, as shown in Table~\ref{tab:mrs}. 
A test case is a couple $(map, traj)$ where $map$ corresponds to a semantic segmentation map and $traj$ is formally handled by a set of waypoints corresponding to the input trajectory of the past motion history.

\begin{table}[t]
\centering
\begin{tabular}{llcc}
    \toprule
    MR & Name & Trajectory & Map \\
    \midrule
    \mrmirror & Mirroring & \cmark & \cmark \\
    \mrrot & Rotating & \cmark & \cmark \\
    \mrscale & Rescaling & \cmark & \cmark \\
    \midrule
    \mrcc & Class Changing & \xmark & \cmark \\
    \mrobs & Obstacle Appearance & \xmark & \cmark \\
    \bottomrule
\end{tabular}
\caption{MRs for multimodal HTP: Applicable input sources. \cmark/\xmark indicate if the input source is modified by the MR.\label{tab:mrs}}
\end{table}

The first group applies basic transformations such as mirroring (\mrmirror), rotating (\mrrot), and rescaling (\mrscale) on the combination of the input trajectory and the semantic segmentation map.  
All of these MRs are revertible, i.e., the source test case can be reconstructed from the follow-up test case, and label-preserving, i.e., if available ground-truth labels were transformed similarly they could be evaluated. However, they require that the coordinate systems for the different inputs are aligned (to ensure spatial consistency) and remain aligned under the transformation, which is usually handled through the preprocessing of the HTP system already.
The HTP system should be robust against each of these transformations and should not change its predictions, i.e., it shall not violate any PVC, namely the WVC (as per Def.~\ref{DefW}) for trajectory distributions and the HVC (as per Def.~\ref{DefH}) for the underlying probability map.

The second group specifically manipulates the semantic segmentation map without modifying the input trajectory.
Possible manipulations are changing the semantic class (\mrcc) or introducing an obstacle on the map (\mrobs).
These MRs are not label-preserving, i.e., we expect a difference in the output of the SUT. Therefore, PVC is not applicable. Instead, we then use HVC, which handles the violation between probability distribution maps.

For all of our MRs, ground-truth labels, i.e., what exactly is the trajectory taken by the human, are not required by \tool.

Table~\ref{tab:mrs} gives a concise overview over all MRs.
In the following, we describe each of the MRs in a structured manner through their inputs, transformation, relation, and parameters.

\subsubsection{Trajectory-related MRs}

\begin{relation}{Mirroring (\mrmirror)}{mirror}
\small
\begin{description}[noitemsep]
    \item[Inputs] Input trajectory, semantic segmentation map
    \item[Transformation] The input is mirrored along the horizontal or vertical axis of the segmentation map.
    \item[Relation] Equivalence relation. Mirroring is a basic transformation, and the HTP model should be robust against it. Mirroring can cause corruption when applied to the original image; for example, in Figure~\ref{fig:input_output} the label ``Slow'' on the street would be unreadable. The segmentation map (Figure~\ref{fig:segmentation}) does not have this level of detail and is not corrupted by the mirroring operation.
    \item[Parameters] Mirror axes: vertical or horizontal
    \item[Violation Criterion] WVC for the trajectory distribution, HVC for the probability map.
\end{description}
\end{relation}

\begin{relation}{Rotation (\mrrot)}{rotation}
\small
\begin{description}[noitemsep]
    \item[Inputs] Input trajectory, semantic segmentation map
    \item[Transformation] The input is rotated by 90/180/270 degrees.
    \item[Relation] Equivalence relation. Rotation is a basic transformation, and the HTP model should be robust against it.
    \item[Parameters] The rotation is limited to multiples of 90 degrees to avoid cutting parts of the segmentation map or introducing background regions in the corners.
    \item[Violation Criterion] WVC for the trajectory distribution, HVC for the probability map.
\end{description}
\end{relation}

\begin{relation}{Rescale (\mrscale)}{rescale}
\small
\begin{description}[noitemsep]
    \item[Inputs] Input trajectory, semantic segmentation map
    \item[Transformation] The rescaling factor of the original image is modified. This MR considers a technical necessity of modern computer vision architectures that inputs must be rescaled to certain sizes, e.g. multiples of 32 to match the setup of the initial convolutional neural network layers. Original input images are resized before being processed. 
    However, the exact input size is not fixed and can be varied.
    \item[Relation] Equivalence relation. The rescaling should, when applied within bounds, not affect the result.
    \item[Parameters] Direction of rescaling, that is, whether to scale up or down, and effect size, that is, how much to change the scaling.
    \item[Violation Criterion] WVC for the trajectory distribution, HVC for the probability map.
\end{description}
\end{relation}

\subsubsection{Map-related MRs}

Map-related MRs target the environmental context of the trajectory. 
They manipulate the semantic map of the surroundings of the human.
We consider two MRs in this category, changing the semantic class of an area in the map and the appearance of an obstacle in the human's predicted trajectory.

\begin{relation}{Semantic Class Change (\mrcc)}{classchange}
\small
\begin{description}[noitemsep]
    \item[Inputs] Semantic segmentation map
    \item[Transformation] Select a transition pair (source class, target class) and replace all occurrences of the source class by the target class.
    \item[Relation] Depending on the transition pair, three relations can occur: (a) the area becomes more walkable, (b) the area becomes less walkable, and (c) the area becomes an obstacle.
    \item[Parameters] List of transition pairs and their effects.
    \item[Violation Criterion] Hypothesis testing criterion on changed cells in the probability map. For a more or less walkable target class, we apply the one-sided test that the probabilities increase, resp. decrease. For an obstacle target class, we test for a decreased probability and check for intersections of the predicted trajectories.
\end{description}
\end{relation}

Table~\ref{tab:class_change_transitions} describes a set of class change transitions for semantic classes, as we use them in the experimental evaluation.
Here, a class change can make the target area either more walkable (expected effect: increased likelihood to be entered by the pedestrian), less walkable (expected effect: decreased likelihood), or render it non-walkable (expected effect: avoidance, the area becomes an obstacle).

\begin{table}[ht]
    \centering
    \begin{tabularx}{\textwidth}{>{\raggedright\arraybackslash}X>{\raggedright\arraybackslash}Xl}
    \toprule
    Original Class & Modified Class & Effect\\\midrule
    Pavement, Terrain & Road & Decrease\\
    Road & Pavement, Terrain & Increase\\
    Road, Pavement, Terrain & Structure, Tree & Avoidance\\
    Structure, Tree & Road, Pavement, Terrain & Increase \\
    Terrain & Pavement & Increase\\
    Pavement & Terrain & Decrease\\
    \bottomrule
    \end{tabularx}
    \caption{\mrcc -- Semantic Class Change: Transitions and their effect on the likelihood to be entered by the pedestrian (Increase/Decrease) or avoidance for a physical obstacle.}
    \label{tab:class_change_transitions}
\end{table}

\begin{figure}[t]
    \centering
     \begin{subfigure}[t]{0.48\columnwidth}
        \centering
        \includegraphics[width=\textwidth]{images/segmentation.png}
        \caption{Image Segmentation Output. Six classes can be annotated.}
        \label{fig:segmentation2}
    \end{subfigure}
    \hfill
    \begin{subfigure}[t]{0.48\columnwidth}
        \centering
        \includegraphics[width=\textwidth]{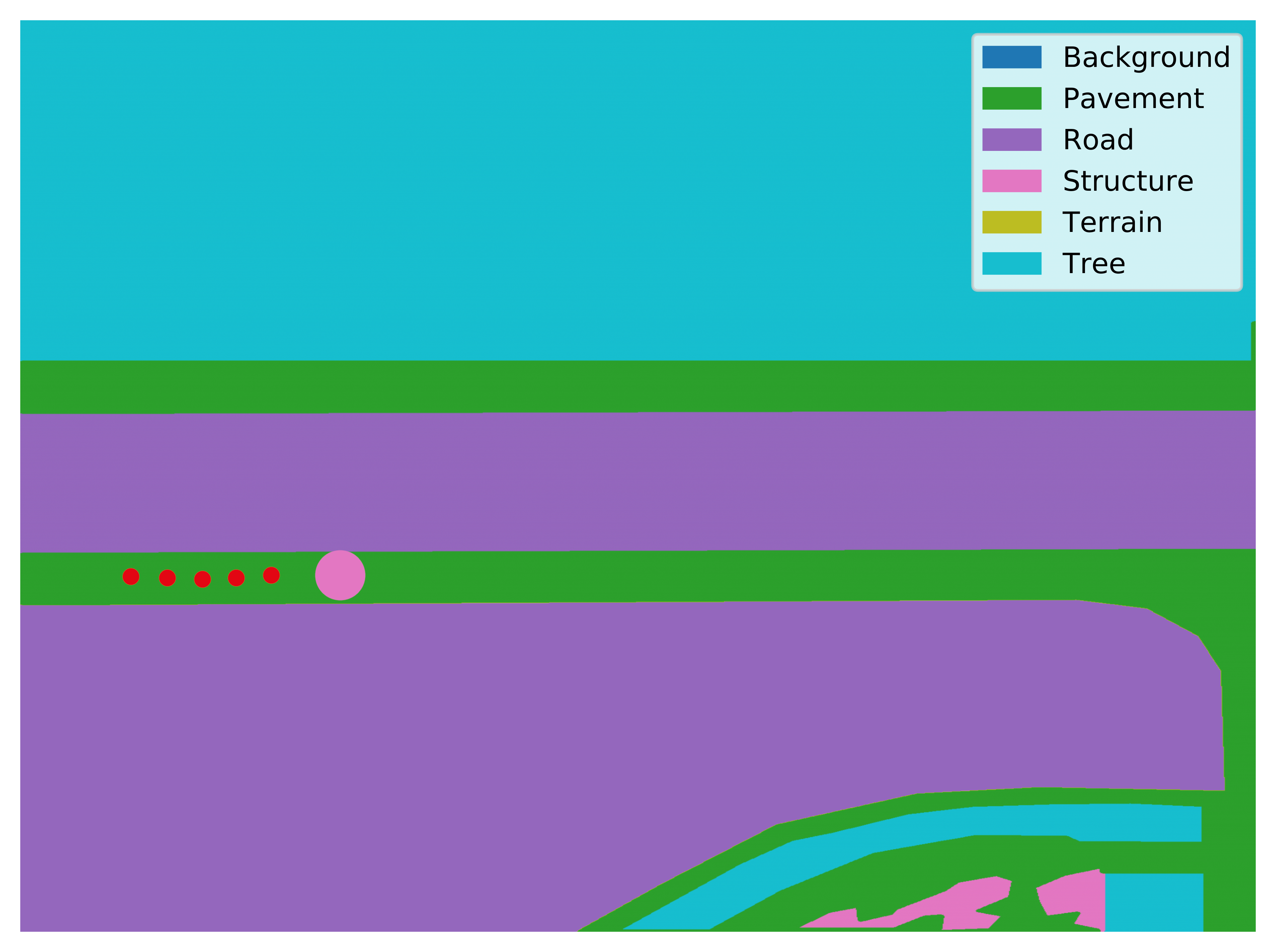}
        \caption{Image Segmentation with two MRs applied.}
        \label{fig:mrapplied}
    \end{subfigure}
    \caption{Example of \mrcc and \mrobs applied simultaneously on a segmentation map. The terrain area at the bottom is changed to road and an obstacle is added in the pedestrian's initially predicted path. Data shows the little\_1 scene from the Stanford Drone Dataset~\citep{robicquet2016learning}.}
    \label{fig:mr_examples}
\end{figure}

As an example, the MR could change the terrain area at the bottom of Figure~\ref{fig:segmentation} into a road (as shown in Figure~\ref{fig:mr_examples}), making it less likely for the human to walk on it. In another case, the grass area might be converted to pavement, which is more likely to happen in another similar area.

\begin{relation}{Obstacle Appearance (\mrobs)}{obstacle}
\small
\begin{description}[noitemsep]
    \item[Inputs] Trajectory predicted by source test case $r$, segmentation map
    \item[Transformation] Given the source test case's output, the follow-up test case is constructed by placing an obstacle in the predicted trajectory.
    In our implementation, we define \textit{structures} and \textit{tree} to be non-passable classes and model the obstacle as a 12-sided polygon to be placed within the predicted trajectory.
    \item[Relation] The expected result is that the follow-up result avoids the obstacle.
    \item[Parameters] The metamorphic relation can be parameterized by the obstacle's class, size, and distance of the obstacle to the human.
    \item[Violation Criterion] We test for a decreased probability on the map and check for intersections of the predicted trajectories, in the same way as for \mrcc with the obstacle target class.
\end{description}
\end{relation}

Figure~\ref{fig:mr_examples} shows an example of the introduction of a class structure obstacle in the path of the pedestrian, which was initially the straight path on the pavement. In the follow-up test case, the HTP system is challenged to predict a trajectory around the obstacle.

\subsection{Test Process}

\begin{algorithm}[th!]
    \small
    \renewcommand{\algorithmicrequire}{\textbf{Input:}}
    \caption{HTP Test Process Overview}\label{alg:cap}
    \begin{algorithmic}[1]
    \Require $HTP\text{: System-under-Test}$
    \State $SourceResults \gets \varnothing$, $ViolationCounter \gets 0$
    \State $S \gets \text{Sample source test case}$ \Comment{Preparation Phase}
    \For{$i \gets 1$ to $N$}
        \State $r \gets HTP.predict(S)$
        \State $SourceResults \gets SourceResults \cup \{r\}$         
    \EndFor
    \State $D_{Src} \gets \text{PairwiseDistances}(f_D, SourceResults)$
    \State $\langle \mu_{Src}, \sigma_{Src} \rangle \gets \text{CalculateVariationMeasures}(D_{Src})$
    \State $MR \gets \text{Select MR to apply}$
    \Comment{MT Phase}
    \State $FU \gets MR.transform(S)$
    \State $R_{FU} \gets HTP.predict(FU)$
    \For{$R_{S} \in SourceResults$} \Comment{Evaluation Phase}
        \State $r \gets SUT(S)$
        \State $R_{S}' \gets MR.transform(R_{S})$
        \State $D \gets \text{WVC}(R_{FU}, R_{S}')$
        \State $PValue = ZTest(D, \mu_{Src}, \sigma_{Src})$
        \If{$PValue \leq 0.05$}
            \State $ViolationCounter = ViolationCounter + 1$
        \EndIf
    \EndFor
    \State \Return $ViolationCounter$
    \end{algorithmic}
\end{algorithm}

Algorithm~\ref{alg:cap} outlines the MT process for a single source and follow-up test case.
The process follows the general structure of the three phases of MT:
First, the source test case is sampled, and the system-under-test is executed with it.
In our case, to handle the non-determinism in the HTP model, we execute the SUT multiple times --- adjustable by the parameter $N$ --- and calculate the pairwise distances between the predictions and calculate statistics.
Afterwards, the test case is transformed according to the selected MR and executed once.
In the evaluation phase, the result of the follow-up test case is compared with each source test case execution, and the z-test is calculated to detect potential violations.

\section{Empirical Evaluation}\label{sec:evaluation}

\subsection{Experimental Setup}

\subsubsection{Datasets}

We use the Stanford Drone Dataset (SDD)~\citep{robicquet2016learning} and the intersection drone dataset (inD)~\citep{Bock2020}, which are known in the trajectory prediction literature~\citep{Mangalam_2021_ICCV,luo2023gsgformer}.
The SDD dataset consists of 11,000 unique pedestrians in eight top-down scenes around the Stanford University campus; the inD dataset consists of 5,300 VRUs (vulnerable road users) at German traffic intersections.
To avoid data leakage, we take the scenes from the test splits of the datasets as in~\citet{Mangalam_2021_ICCV}.

Since we utilize the existing test sets, we have ground-truth information available for our experiments.
We use this ground-truth information to calculate standard trajectory prediction metrics for the source and follow-up predictions. These metrics form a reference for interpreting the effectiveness of the stochastic violation criterion and the general effect of metamorphic transformations on prediction performance.

\subsubsection{HTP Models}

The system under test (SUT) is the Y-net trajectory prediction model~\citep{Mangalam_2021_ICCV} using the publicly available trained model weights\footnote{Online: \url{https://github.com/HarshayuGirase/Human-Path-Prediction}} and the experimental parameters.
Most experimental parameters follow the settings used in the Y-net experiments to maintain their reported quality. When deviating or introducing new parameters, we mention the intention for their values specifically.
We tested trajectory prediction in the short-term setting (SDD) with $t_p = 3.2$ second past motion history, sampled at 2.5 FPS, and a prediction horizon of $t_f = 4.8$ seconds.
In long-term forecasting (SDD and inD) it is $t_p = 5$ second past motion history, sampled at 1 FPS, and a prediction horizon of $t_f = 30$ seconds. 
In all settings, Y-net samples $K=20$ trajectories per prediction.

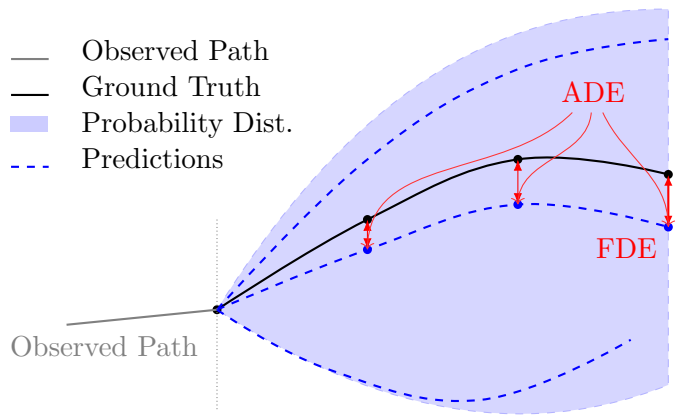
\begin{figure}
    \centering
    \begin{tikzpicture}[
  gt/.style={black, thick},
  pred/.style={blue, thick, dashed},
  obs/.style={gray, thick},
  err_line/.style={red, <->, {Latex[length=1.5mm]}-{Latex[length=1.5mm]}},
  point/.style={circle, fill, inner sep=1.2pt},
  label/.style={font=\small, align=left},
]
\begin{pgfonlayer}{background}
  \fill[blue!20, fill opacity=0.8, draw=blue!30, dashed]
    (0,0)
    .. controls (2,3.2) and (4,4.0) .. (6,4.0)
    -- 
    (6,-1.0)
    .. controls (4,-1.8) and (2,-1.3) .. (0,0)
    -- cycle;
\end{pgfonlayer}

\node[point, black] at (0,0) {};

\draw[obs] (-2,-0.2) -- (0,0);
\node[label, color=gray, below] at (-1.5, -0.2) {Observed Path};

\draw[gt] (0,0) plot [smooth, tension=0.7] coordinates { (0,0) (2,1.2) (4,2.0) (6,1.8) };
\foreach \x/\y in {2/1.2, 4/2.0, 6/1.8} {
  \node[point, black] at (\x,\y) {};
}
\coordinate (gt1) at (2,1.2);
\coordinate (gt2) at (4,2.0);
\coordinate (gt3) at (6,1.8);

\draw[pred] (0,0) plot [smooth, tension=0.7] coordinates { (0,0) (2,0.8) (4,1.4) (6,1.1) };
\foreach \x/\y in {2/0.8, 4/1.4, 6/1.1} {
  \node[point, blue] at (\x,\y) {};
}
\coordinate (p1) at (2,0.8);
\coordinate (p2) at (4,1.4);
\coordinate (p3) at (6,1.1);

\draw[pred] (0,0) plot [smooth, tension=0.7] coordinates { (0,0) (2,2.0) (4,3.2) (6,3.6) };
\draw[pred] (0,0) plot [smooth, tension=0.7] coordinates { (0,0) (1.5,-0.8) (3.5,-1.2) (5.5,-0.4) };

\draw[err_line] (gt1) -- (p1);
\draw[err_line] (gt2) -- (p2);
\draw[err_line, thick] (gt3) -- (p3) node[below left, label, color=red] {FDE};

\node[label, color=red] at (5.0, 2.9) (ade_label) {ADE};
\draw[->, red!80, thin] (ade_label) to[out=-65, in=90] (p3);
\draw[->, red!80, thin] (ade_label) to[out=-95, in=90] (p2);
\draw[->, red!80, thin] (ade_label) to[out=-135, in=90] (p1);

\draw[gray, densely dotted] (0,1.2) -- (0,-1.35); %

\node[align=left, label, font=\small, inner sep=3pt] at (-0.8, 2.7) {
  \begin{tabular}{@{}ll@{}}
  \tikz\draw[obs] (0,0.05) -- (0.5,0.05); & Observed Path \\
  \tikz\draw[gt] (0,0.05) -- (0.5,0.05); & Ground Truth \\
  \tikz\fill[blue!20] (0,0) rectangle (0.5,0.2); & Probability Dist. \\
  \tikz\draw[pred] (0,0) -- (0.5,0); & Predictions \\
  \end{tabular}
};
\end{tikzpicture}
    \caption{Best-of-N ADE and FDE: The model generates N plausible future paths (here, N=3) from a probability distribution (blue-shaded background, simplified). In Best-of-N, the path with the minimum error is chosen to calculate ADE/FDE. ADE is the average of the all red lines. FDE is the length of only the thick red line.}
    \label{fig:adefde}
\end{figure}

Per source test case, we sample $N=8$ sets of solutions to calculate the violation threshold and compare the follow-up test cases against it.
The value is chosen from preliminary experiments to establish a sufficient basis for the selection of the violation threshold while avoiding unnecessary computational cost.
We report the violation rate, i.e., the percentage of prediction comparisons for which the distance exceeds the threshold, as the main metamorphic testing criterion.
We further calculate the average performance of the source and follow-up test in terms of average (ADE) and final displacement error (FDE), the standard evaluation metrics for HTP:
\begin{align}
    ADE &= \frac{1}{N \times T_p} \sum_{n \in N} \sum_{t \in T_p} \lVert \hat{p}_t^n - p_t^n \rVert_{2}\\
    FDE &= \frac{1}{N} \sum_{n \in N} \lVert \hat{p}_{T_p}^n - p_{T_p}^n \rVert_{2}
\end{align}
ADE is the average distance between the prediction and the ground-truth trajectory, meaning it measures how close the two paths are to each other. 
FDE is the distance between the trajectory endpoints, that is, it measures only how close the endpoints are, independent of the paths leading up to them. 
The common evaluation setup is Best-of-N (BoN), which means that the smallest ADE and FDE are reported over N sampled trajectories, i.e., $K=20$ in our experiments.
We visualize Best-of-N ADE/FDE in Figure~\ref{fig:adefde}.
Since BoN evaluation does not consider the distribution of the trajectories in addition to the best one, we additionally calculate the mean ADE and FDE over all predicted paths.
To identify MR violations, we apply a similar approach to the WVC and compare the ADE/FDE of the follow-up test case to the averaged results of all sampled source test cases via a z-test.
We denote the two sets of metrics as \textit{BoN-ADE}, \textit{BoN-FDE}, \textit{Mean-ADE}, and \textit{Mean-FDE}.
These four metrics require ground-truth information, which is generally not available in MT. They are included in the experiments to evaluate the utility of the MRs and the WVC.

For the MR Rescale, we choose two different rescale values $0.2$ and $0.3$, which slightly deviate from the Y-net default value of $0.25$.
These values are picked since they are close to the default value and should not introduce a too strong distribution shift for the model, but still cause the model input to be differently sized after all preprocessing steps, and therefore test the initial value's robustness.

\subsubsection{Technical Setup}\label{sec:setup}

Our implementation is based on the Y-net codebase and uses POT (Python Optimal Transport) to calculate Wasserstein distances~\citep{flamary2021pot}.
The source code for our experiments and the experimental results are available in our replication package\footnote{Replication package: \url{https://zenodo.org/records/15862940}}.

\subsection{Results}

We structure the discussion of the results along the two groups of MRs.

\subsubsection{Label-preserving Metamorphic Relations}

Table~\ref{tab:results_traj} lists the results for the label-preserving MRs \mrmirror, \mrrot, and \mrscale.
We observe a close similarity in detected violations of the metamorphic relation for the proposed Wasserstein violation criterion, which does not need any ground-truth labels, and the ground-truth-dependent Mean-ADE and Mean-FDE.
This similarity occurs in all settings and datasets.
There is also a strong difference in ADE/FDE values between BoN and Mean.

In addition to the violation rate based on the WVC, we show the average Hellinger distance over all source and follow-up test cases for an MR.
We report these distance values to highlight the differences in the degree to which MRs affect the resulting probability map.
The smallest difference is observed for the 180-degree rotation; mirroring the input has a larger effect but is independent of the mirror axis, whereas 90/270-degree rotations and rescaling of the input have the largest effect.
A strong difference between short-term predictions and long-term predictions (both on SDD and inD) is visible, too.

There are diverse aspects to the interpretations of these results:
One aspect is that even though we manipulate only the segmentation map, which should not contain specific environment features such as text, there are still some observable patterns in the segmentation map and then the transformation could cause a distribution shift, which the model cannot handle. 
These patterns could also occur in the behaviour of the pedestrians in the data set.
It is therefore important to consider whether the sensitivity of the model represents a failure of the Y-net architecture to achieve geometric invariance, or if the model has correctly learned patterns of pedestrian behaviour present in the training data. 
The SDD and inD datasets are filmed from fixed top-down perspectives where traffic flow, pedestrian crossings, and building layouts are not rotationally symmetric.  In this context, a 'violation' of the rotation MR does not necessarily indicate a bug, but rather successfully reveals that the model has learned a strong environmental prior. This highlights a crucial function of \tool: not just finding faults, but characterizing the implicit assumptions a model has learned.

In another aspect, rotating 90 and 270 degrees causes flipping the aspect ratio of the segmentation map, again causing a distribution shift over the training data, which are mostly in landscape orientation.
Since all these MRs are label-preserving, we would only expect minimal changes and for all MR groups distances of similar magnitude, whereas the results shown here indicate some robustness issues in the waypoint map prediction of the system.
Finally, the longer forecasting period allows more variability and a larger accumulation of deviations, therefore also leading to a larger overall difference. This causes an increase in the HVC, which is not normalized or standardized over the forecasting period, but self-adapts over the preparation phase of the MT testing procedure (see Algorithm~\ref{alg:cap}).

\begin{table}[thp]
\adjustbox{max width=\textwidth}{%
    \centering
    \footnotesize
    \renewcommand{\arraystretch}{0.9}
    \begin{tabular}{ll|ccccc|c}
\toprule
           D &          MR &   WVC &  B-ADE &  B-FDE &  M-ADE &  M-FDE &        HVC \\
\midrule
 \parbox[t]{2mm}{\multirow{7}{*}{\rotatebox[origin=c]{90}{SDD (Short)}}} &    Mirror-v & 61.7 &  27.1 &  26.1 &  65.1 &  63.4 &  0.68±0.19 \\
  &    Mirror-h & 61.6 &  26.2 &  26.9 &  64.8 &  63.2 &  0.72±0.15 \\
\cmidrule(lr){2-8}
  &   Rotate-90 & 82.3 &  41.7 &  35.9 &  83.2 &  83.0 &  0.79±0.21 \\
  &  Rotate-180 & 81.0 &  43.4 &  37.0 &  83.2 &  83.1 &  0.26±0.09 \\
  &  Rotate-270 & 84.8 &  39.3 &  35.1 &  83.9 &  84.2 &  0.96±0.24 \\
\cmidrule(lr){2-8}
  &  Resize-0.2 & 71.3 &  39.8 &  32.0 &  75.3 &  74.4 &  1.05±0.08 \\
  &  Resize-0.3 & 70.1 &  30.9 &  27.4 &  74.6 &  72.0 &  0.94±0.08 \\
  \midrule
\parbox[t]{2mm}{\multirow{7}{*}{\rotatebox[origin=c]{90}{SDD (Long)}}} &    Mirror-v & 36.4 &  41.1 &  35.1 &  39.3 &  34.0 &  4.22±1.07 \\
   &    Mirror-h & 31.0 &  44.0 &  33.5 &  36.6 &  34.3 &  4.38±0.95 \\
 \cmidrule(lr){2-8}
  &   Rotate-90 & 52.6 &  44.5 &  32.7 &  47.9 &  44.8 &  5.97±1.69 \\
   &  Rotate-180 & 52.3 &  46.1 &  37.2 &  49.2 &  44.0 &  2.82±0.82 \\
   &  Rotate-270 & 51.1 &  47.1 &  37.2 &  50.0 &  44.5 &  7.47±2.11 \\
 \cmidrule(lr){2-8}
  &  Resize-0.2 & 38.7 &  45.8 &  33.2 &  46.6 &  41.4 &  7.61±1.32 \\
   &  Resize-0.3 & 49.8 &  41.4 &  34.0 &  50.5 &  44.8 &  7.44±1.22 \\
  \midrule
\parbox[t]{2mm}{\multirow{7}{*}{\rotatebox[origin=c]{90}{inD (Long)}}} &    Mirror-v & 59.7 &  51.7 &  49.4 &  71.8 &  73.6 &  5.55±0.80 \\
   &    Mirror-h & 62.7 &  58.0 &  44.3 &  66.1 &  65.5 &  4.91±0.76 \\
 \cmidrule(lr){2-8}
  &   Rotate-90 & 93.1 &  62.6 &  46.0 &  81.0 &  83.9 &  6.74±1.01 \\
   &  Rotate-180 & 93.5 &  62.1 &  43.1 &  75.3 &  93.1 &  2.80±0.87 \\
   &  Rotate-270 & 74.3 &  56.3 &  36.2 &  77.0 &  77.0 &  5.33±0.95 \\
 \cmidrule(lr){2-8}
  &  Resize-0.2 & 79.2 &  74.1 &  60.9 &  79.3 &  73.6 &  5.48±1.49 \\
   &  Resize-0.3 & 64.4 &  56.9 &  42.0 &  66.1 &  68.4 &  6.27±0.90 \\
\bottomrule
\end{tabular}
}
    \caption{Violation rates (in \%) per label-preserving metamorphic relation and compared to the labelled baselines. For HVC, we report the mean distance + std. dev. between source and follow-up test case. D: Dataset; BoN: Best-of-N.}
    \label{tab:results_traj}
\end{table}

\paragraph{Agreement of WVC violations and ADE/FDE} We perform an additional experiment to investigate the agreement between the violations detected by WVC and the criteria based on ADE/FDE.
The experiment is approached as a binary classification problem, where ADE/FDE-detected violations are considered class labels, and WVC-detected violations are predictions. We also report accuracy, precision, and recall over multiple p-value thresholds, i.e., over which p-value is a result identified as a violation, to understand the sensitivity of the results.

\begin{figure}[t]
    \centering
    \includegraphics[width=\textwidth]{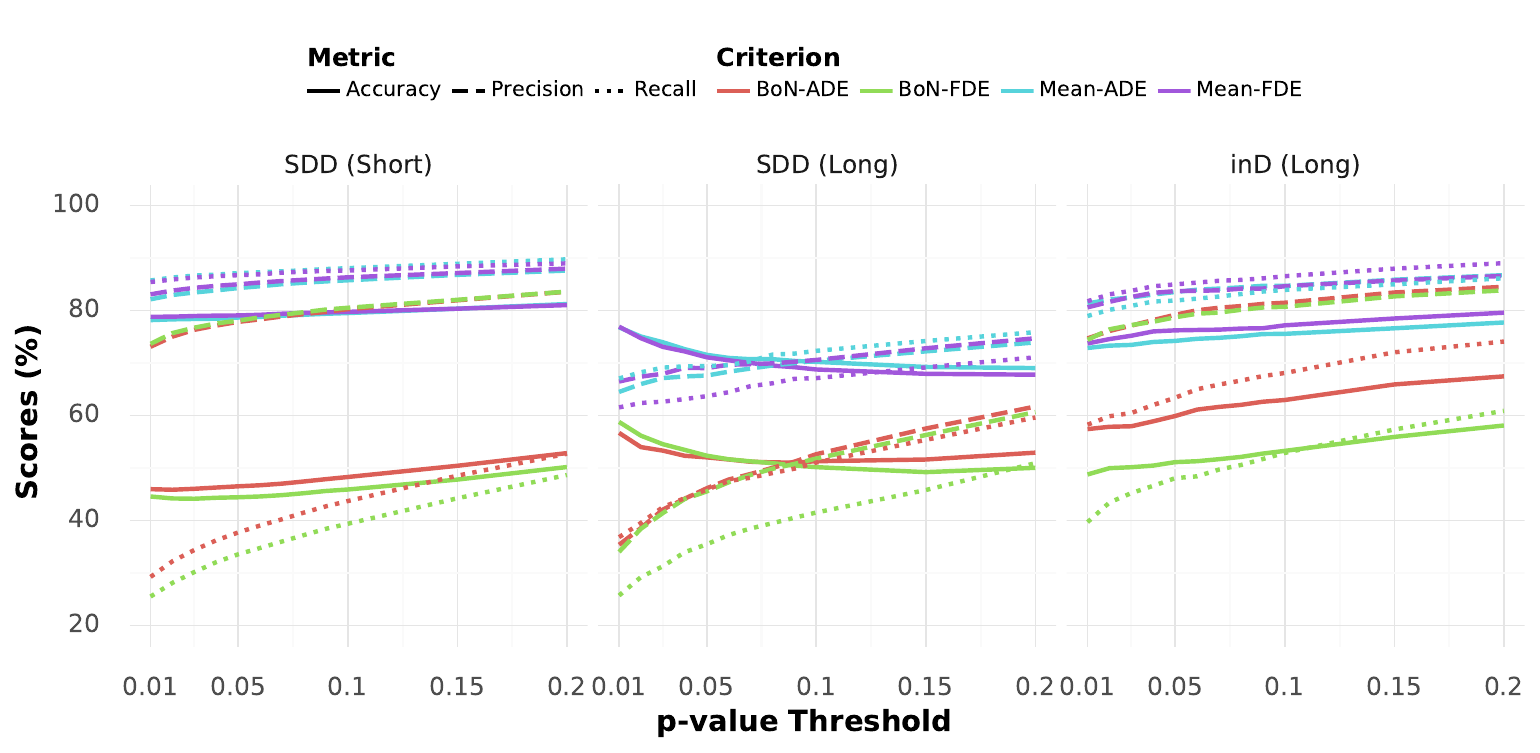}
    \caption{Agreement of WVC violations and ADE/FDE: Dependency between p-value threshold and classification scores for label-preserving MRs; results are aggregated over \mrmirror, \mrrot, and \mrscale.}
    \label{fig:scores_shortterm}
\end{figure}

The results are shown in Figure~\ref{fig:scores_shortterm} for the three experimental settings.
They confirm that there is substantial agreement between the detected violations of WVC and Mean-ADE/FDE.
At the same time, they show that the p-value threshold is relevant to be adjusted, even though with a moderate effect only.

\subsubsection{Map-oriented Metamorphic Relations}

The results for MRs that manipulate the segmentation map, \mrcc and \mrobs, are shown in Table~\ref{tab:results_map}.
First, we observe a drastic difference between the types of class changes.
Class changes that should increase the likelihood for a pedestrian to access them never lead to an MR violation, i.e., the probability of these areas was never significantly reduced, even though the Hellinger distance on the probabilities of the changed area is the highest in this case.
For class changes that should decrease the likelihood that a pedestrian can access them, we observe the opposite effect, that is, a high number of MR violations.
Again, there are multiple interpretation approaches to this observation:
One aspect is that the input trajectory usually is in an area that has a high probability of being walked on, and the immediate future areas are commonly also of high probability. When this area is now randomly changed, the probability map prediction still assigns some amount of probability to these areas due to the spatial proximity to the pedestrian.
Another aspect lies in the classification of segmentation classes that have a higher or lower probability of walking (see Table~\ref{tab:class_change_transitions}), which is picked manually and could be adopted depending on the exact test conditions.

We separately list the results for \mrobs and \mrcc with an obstacle effect, i.e. the area becoming a structure or tree. 
The class change is less effective in leading to violations of the MR, but causes a higher number of intersections of the predicted trajectories with obstacle areas.
Adding obstacles to the initially predicted path of the pedestrian, as does \mrobs, is more effective in terms of probability changes, but does not cause the model to predict that the pedestrian will walk through the obstacle.
This is a positive result for the HTP system, as it is capable of sufficiently recognizing and avoiding obstacles in close proximity, even if they appear on the originally predicted future trajectory.

\begin{table}[th]
    \centering
    \small
    \begin{tabular}{lll|rrr}
\toprule
           D &            MR &    Effect &    HTC &  Intersections &        HVC \\
\midrule
 \parbox[t]{2mm}{\multirow{4}{*}{\rotatebox[origin=c]{90}{SDD (Short)}}} &  Class Change &  Increase &  97.9 &           -- &  0.10±0.14 \\
  &  Class Change &  Decrease &   0.0 &           -- &  0.18±0.24 \\
  &  Class Change &  Obstacle &   3.5 &          27.7 &  0.12±0.20 \\
  &      Obstacle &  Obstacle &  22.9 &           9.2 &  0.01±0.01 \\
  \midrule
\parbox[t]{2mm}{\multirow{4}{*}{\rotatebox[origin=c]{90}{SDD (Long)}}} &  Class Change &  Increase &  94.5 &           -- &  2.84±2.55 \\
   &  Class Change &  Decrease &   8.9 &           -- &  5.75±3.10 \\
   &  Class Change &  Obstacle &  10.7 &          33.4 &  3.74±3.49 \\
   &      Obstacle &  Obstacle &  39.0 &           1.6 &  0.13±0.11 \\
  \midrule
\parbox[t]{2mm}{\multirow{4}{*}{\rotatebox[origin=c]{90}{inD (Long)}}} &  Class Change &  Increase &  93.1 &           -- &  1.85±1.94 \\
   &  Class Change &  Decrease & 100.0 &           -- &  3.67±1.02 \\
   &  Class Change &  Obstacle &  61.5 &          46.9 &  3.93±0.96 \\
   &      Obstacle &  Obstacle &  48.9 &           9.7 &  0.37±0.29 \\
\bottomrule
\end{tabular}

    \caption{Violation rate, intersections and Hellinger distance for map-oriented MRs. For HVC, we report the violation rate (in \%). For HVC, we report the mean distance + standard deviation between source and follow-up test case. D: Dataset; HTC: Hypothesis Testing Criterion; HVC: Hellinger Violation Criterion.}
    \label{tab:results_map}
\end{table}

For these MRs, we observe that the p-values are either very small or close to one; therefore, we do perform an evaluation of the effect of the p-value threshold, unlike in the previous result section.

\section{Related Work}\label{sec:rel_work}

\paragraph{Testing HTP}
Forecasting the trajectory of pedestrians based on their past movements is important to design safe automated driving systems. Previous work has addressed the challenge of verifying the robustness of HTP models by considering adversarial attacks~\citep{Zhang_2022_CVPR,cao2022advdo,pmlr-v205-cao23a,Zheng_2023_WACV,pmlr-v211-tan23a,jiao2022semi}. 
However, many of these works have just translated adversarial attacks proposed in the context of image classification and object detection tasks without taking into account the peculiarities of HTP model robustness verification. Recently, using Probably Approximately Correct learning (PAC) and formalizing the notion of HTP robustness, \citeauthor{Zhang_2023_ICCV} has proposed in \citep{Zhang_2023_ICCV} a rich framework to verify the robustness of pedestrian trajectory prediction models. Using ablation studies, \citeauthor{Uhlemann2024} have proposed evaluating the safety of the HTP model in the context of automated driving~\citep{Uhlemann2024}. 

\paragraph{Statistical MT}
To our knowledge, MT has not yet been used to test HTP models, but approaching the verification of stochastic systems with MT is not new \citep{Chen2018,olsen_increasing_2019}. Introduced by \citet{guderlei2007statistical}, statistical MT replaces traditional violation criteria, i.e., the detection of MR violated, with hypothesis testing. 
Used for testing statistical optimization algorithms, for example, simulated annealing, statistical MT reveals itself to be interesting but also dependent on the problem to be solved with respect to its performance \citep{Yoo2010}. 
We believe that this approach, i.e., statistical MT, is relevant to test HTP models and to adopt it accordingly for \tool. We apply probabilistic violation criteria directly to the probability distributions returned by the system, as opposed to testing its stochastic outputs over many sampling runs.
Recently, a different perspective on statistical MT has been proposed, such that -- using statistical techniques -- the suspiciousness of each test case is estimated first to improve MT's efficiency \citep{zheng_identifying_2025}.

\paragraph{MT for Image Processing}
At the same time, several studies applied MT to image processing models, similar to the semantic image segmentation model in the overall HTP system. The range of application and testing purposes is broad as there are many facets in image processing, ranging from the model implementation itself, over the validation of learned weights, to ways on how to integrate MT to improve the model performance at test time or deployment.
One of the first studies to address image processing is by \citet{guderlei_towards_2007}, albeit to test traditional image processing software, not learned image processing models.
\citet{Spieker2020} consider image classification and object detection as case studies to learn robust boundaries for different parametrized MRs. 
Similarly, \citet{torikoshi_sensitive_2023} test image classification models, while guiding the metamorphic transformation from explainable AI techniques that provide information about the relevance of each individual pixel in the image.
A framework to generate new test inputs through generative AI techniques was demonstrated by \citet{sun_metamorphic_2024}.
\citet{dwarakanath_identifying_2018} present MT for finding implementation bugs in image classifiers, using a variation of MRs like changing the image colour channels or, similar to our work, rescaling the test data.
Within larger scientific application development, \citet{ding_validating_2017} deploy iterative MT for the validation of a 3D structure reconstruction software of mitochondria in cells.
Other work considers the use of MRs to enhance machine learning classifiers \citep{Xu2018,DBLP:journals/jss/XuTFBZC21} or to detect adversarial examples on these models through affine transformations (rotation, shearing, scaling, translation) \citep{mekala_metamorphic_2019}.

\section{Threats to Validity}

There are some validity threats that must be mentioned to place the study and its result in an appropriate context.
We consider only a subset of possible transformations of the input sources.
Our selection of MRs is effective, there are other MRs possible, e.g. related to time dilation, trajectory inversion, or the entire construction of segmentation maps. Our MRs do not cover all features, and other MRs might be necessary to be exhaustive.

In the design of the probabilistic violation criterion, we make a decision for distance metrics, i.e., Wasserstein and Hellinger, and statistical tests, i.e., Wilcoxon signed-rank. These are not the only options, and other distance metrics exist; for example, Wasserstein might be replaced with the maximum mean discrepancy or Hellinger with KL-divergence. 
We do not claim that our selection is optimal but reasonable. 
We further argue that making an optimal selection has several influencing parameters, and it is out-of-scope for this paper to perform an extensive study on the adequacy of probability distance metrics. However, we recommend considering selecting alternative distance metrics when implementing \tool in other use cases.

There is a bias from the selection of the HTP system that we apply as SUT, Y-net.
Any effect size of the results can be different in other systems and should not be taken as a generalized statement.
However, we see Y-net as a representative system for our study to discuss the application of metamorphic testing to human trajectory prediction.
The parameterization of the experiments in terms of sampled future trajectories and number of goals follows directly the Y-net configuration, making it closer to the original domain of the SUT, while it might bias the results we observe regarding the reliability of the PVC.

The violation of MRs in our testing approach is based on statistical tests and is, to some extent, subject to stochastic influences. We try to mitigate this risk by selecting appropriate tests and distances to specifically handle the stochastic nature of HTP, but it is not possible to fully encapsulate all randomness and have an entirely deterministic testing procedure. 
However, as long as we can identify any input that causes a MR violation, we can identify a weakness in the HTP system, and the absence of inputs that cause MR violations does not mean that there are none, which is an inherent property of metamorphic testing already.

Our results show a correlation between WVC violations and Mean-ADE/FDE violations. 
While encouraging, this does not prove that WVC is capturing the same underlying flaws, only that the violations tend to co-occur in this specific experimental setup.

An MR violation indicates an inconsistency according to the defined relation. It does not automatically imply a safety-critical failure. A system might violate rotation invariance, but still perform safely in its operational design domain. The link between specific MR violations and actual safety risks needs careful interpretation and additional future work.

Finally, the experimental evaluation is based on our own implementation (available online, see Section~\ref{sec:setup}), using external software libraries and the Y-net source code released by \citeauthor{Mangalam_2021_ICCV}. 
Although we checked our code carefully, there is the risk of faults in our own code that could affect the experimental results.

\section{Conclusion}\label{sec:conclusion}

In this work, we addressed the challenge of testing multimodal Human Trajectory Prediction (HTP) systems, whose stochastic nature and reliance on complex inputs make traditional oracle-based testing difficult. 
We introduced \tool, a framework that uses metamorphic testing specifically adapted for this domain. 
Our primary contribution lies in the development of five domain-specific Metamorphic Relations (MRs) targeting both geometric invariances (mirroring, rotation, scaling of trajectory and map inputs) and semantic map context manipulations (class changes, obstacle insertion).

Critically, to handle the stochastic HTP outputs, we proposed and evaluated probabilistic violation criteria. 
The Wasserstein Violation Criterion (WVC) effectively assesses the equivalence of predicted trajectory distributions for label-preserving MRs, showing a strong correlation with ground-truth-based metrics in our experiments without requiring the ground truth itself. 
Furthermore, the Hellinger Violation Criterion (HVC) provides insights into changes in intermediate probability maps, while the Hypothesis Testing Criterion (HTC) successfully verifies expected directional changes in output probabilities for map-altering MRs designed to induce specific behavioural shifts, e.g., avoidance.

Our empirical evaluation of the Y-net model demonstrated the practical applicability of \tool. 
The framework successfully identified statistically significant deviations from expected behaviour under various transformations, highlighting potential robustness issues, and validating the sensitivity of the proposed violation criteria. 
Notably, the WVC provides a viable oracle-less method for assessing prediction consistency, while HTC confirms the model's response to environmental changes such as obstacles. 
This study thus establishes MT as a valuable and systematic approach to HTP testing, offering a structured, systematic, and oracle-less methodology to improve the robustness and reliability assessment of these critical components in autonomous systems.

In future work, we will expand the applicability of \tool as a general HTP testing tool that can be easily integrated into the HTP training and evaluation process. Currently, this is challenging, since there are commonly used datasets, but most methods apply custom preprocessing and data formats, and there are no commonly used interfaces. To gather adoption, \tool will need to be flexible enough to be called from the HTP system rather than instrument the HTP system. It must also address all components of the HTP pipeline end-to-end, including subsystems, like the semantic segmentation subsystem, which we have excluded for the current version of this study. 
Additionally, we will further consider the modelling of dedicated scenarios via custom segmentation maps and input trajectories for a broader diversity in the scenarios, as well as adaptive parametrization of the metamorphic relations~\citep{Spieker2020} to identify the robustness boundaries of the HTP system. This should support further automation of the metamorphic testing process.

\section*{Acknowledgments}
This work is funded by the European Commission through the AI4CCAM project (Trustworthy AI for Connected, Cooperative Automated Mobility) under grant agreement No 101076911 and by the AutoCSP project of the Research Council of Norway, grant number 324674.

\bibliographystyle{plainnat}
\bibliography{refs}

\end{document}